\documentclass{article}
\usepackage{epsfig}
\usepackage{framed}

\usepackage{amsmath}
\usepackage{amssymb}
\usepackage{amsbsy,todonotes,cancel, mathrsfs}
\usepackage[colorlinks=true, pdfstartview=FitV, linkcolor=blue, citecolor=blue, urlcolor=blue]{hyperref}
\def \scr{\mathscr}
\def \div{{\rm div}\,}
\def \h{{\mathbf h}}
\def \v {{\mathbf v}}
\def \c{{\mathbf c}}
\def\PhiPhi#1#2#3#4#5#6 { { 
 \Cauchy_{#4#2}(#6,#3) \Phi_{#1#5}(#3)
 -\Cauchy_{#1#5}(#3,#6)\Phi_{#4#2}(#6)
 +\big(\Cauchy(#3,#6)\Phi(#6)\big)_{#1#5} \delta_{#4#2}
- \big(\Cauchy(#6,#3)\Phi(#3)\big)_{#4#2} \delta_{#1#5} 
} }

\def\NN#1#2#3#4#5 { { N^{(#1)}_{#2#5} \delta_{#4#3} - N^{(#1)}_{#4#3}\delta_{#2#5} } }

\def \Phikappa #1#2#3#4#5 
{{
\res{#5=t_{#4}} \le(\frac{\Cauchy(#3,#5)}{\d #3_{#4}}   {\d}\frac {\Phi(#5)}{\d #5_{#4}}\ri)_{#1#2}
}}
\def\PhiK#1#2#3#4#5#6#7{ { 
  \Cauchy_{#1#5}(#3,#7) \Cauchy_{#4#2}(#6,#3) 
 + \delta_{#4#2} \big(\Cauchy(#3,#6)\Cauchy(#6,#7)\big)_{#1#5}
 -\Cauchy_{#1#5}(#3,#7) \Cauchy_{#4#2}(#6,#7) 
} }

\def \PhiN #1#2#3#4#5#6
{ {
 \res{#6=t_j} \le(\Cauchy(#3,#6)\Phi(#6)\ri)_{#1#5} \delta_{#4#2}
 - \res{#6=t_j} \Big(\Cauchy_{#1#5}(#3,#6)\Phi_{#4#2}(#6)\Big)
}
}

\def\PhiPhiten #1#2#3#4 {
 \le[\m{\Phi}^{#3}(#4), \m{\Cauchy}^{#1}(#2,#4) \m{\Pi}^{#1#3} \ri] -  \le[\m{\Phi}^{#1}(#2), \m{\Cauchy}^{#3}(#4,#2) \m{\Pi}^{#1#3} \ri] }

\def\PhiKten #1#2#3#4#5 {
\m{\Cauchy}^{#1}(#2,#5)\m{\Cauchy}^{#3}(#4,#5) \m{\Pi}^{#1#3} - \m{\Cauchy}^{#1}(#2,#5)\m{\Cauchy}^{#3}(
#4,#2) \m{\Pi}^{#1#3} - \m{\Cauchy}^{#1}(#2,#4)\m{\Cauchy}^{#1}(#4,#5) \m{\Pi}^{#1#3}
}

\def\TR #1#2#3#4 { 
 \m{\Cauchy}^{#1}(#3,#4) \m{\Pi}^{#1#2} }

\def \Phizeta #1#2#3#4#5
{
{
\res{#5=t_{#4}} \Cauchy_{#1#2}(#3,#5) \d #5_{#4}
}
}

\def\KN #1#2#3#4#5#6
{{
 \res{w=t_j}\Cauchy_{#1#6}(#3,w) \Cauchy_{#5#2}(w,#4) 
}}
\renewcommand{\theequation}{\arabic{section}.\arabic{equation}}

\def \f{ {\mathbf f}}
\makeatletter
\@addtoreset{equation}{section}
\makeatother

\def\Cauchy{\mathbf K}

  \def\nn{\nonumber}
  \newtheorem{theorem}{Theorem}[section]

\newtheorem{proposition}[theorem]{Proposition}
\newtheorem{corollary}[theorem]{Corollary}
\newtheorem{definition}[theorem]{Definition}
\newtheorem{lemma}[theorem]{Lemma}

\newtheorem{example}[theorem]{Example}
\newtheorem{exercise}[theorem]{Exercise}
\newtheorem{examps}[theorem]{Examples}

\newtheorem{remark}[theorem]{Remark}

\def\le{\left}
\def\eqref#1{(\ref{#1})}
\def\ri{\right}

\def\m{\ds \mathop}

\def \QED{\hfill $\blacksquare$\par \vskip 5pt}
\def\ds{\displaystyle}

\def\res{\mathop{\mathrm {res}}\limits_}

\def\br{\begin{remark}\rm}
\def\er{\hfill $\blacktriangle$\end{remark}}
\definecolor{shadecolor}{rgb}{0.95, 0.95, 0.86}
\def\bt{
\begin{shaded}
\begin{theorem}}
\def\et{\end{theorem}
\end{shaded}}
\def\bd{
\begin{shaded}
\begin{definition}}
\def\ed{\end{definition}
\end{shaded}}
\def\bp{
\begin{shaded}
\begin{proposition}}
\def\ep{\end{proposition}
\end{shaded}}

\def\bc{\begin{corollary}}
\def\ec{\end{corollary}}

\def\brs{\begin{remarks}
\begin{enumerate}}

\def\ers{\end{enumerate}\end{remarks}}
\def\bx{\begin{example}\small}
\def\ex{\end{example}}
\def\bxr{\begin{exercise}\small}
\def\exr{\end{exercise}}
\def\bl{\begin{lemma}}
\def\el{\end{lemma}}
\def\bxs{\begin{examps}. \rm\begin{enumerate}}
\def\exs{\end{enumerate}\end{examps}}

\def\be{\begin{equation}}
\def\ee{\end{equation}}

\def\&{\hspace{-15pt}&}
\def\bea#1\eea{\begin{align}#1\end{align}}
\def\beas{\begin{eqnarray*}}
\def\eeas{\end{eqnarray*}}
\def \pa{\partial}
\def\C{{\mathbb C}}

\def\N{{\mathbb N}}

\def\wh{\widehat}
\def\H{{\cal H}}
\def\Z{{\mathbb Z}}

\def\d{\mathrm d}
\def \Mat{{\rm Mat}}
\def\K{\mathcal K}
\def\l{\lambda}

\def\1{{\bf 1}}

\def\wt{\widetilde}
\def\ds{\displaystyle}
\def\e{\mathbf e}
\def\tr{\mathrm {Tr}}
\def\CC{\mathcal C}

\date{}
\begin{document}
\baselineskip 16pt plus 1pt minus 1pt
\vspace{0.2cm}
\begin{center}
\begin{Large}
\textbf{The $r$-matrix structure on the moduli space of framed Higgs pairs}
\end{Large}\\
\bigskip
\begin{large} {M.
Bertola}$^{\ddagger}$\footnote{Marco.Bertola@concordia.ca\\
{\scriptsize Compiled: \today}
}
\end{large}
\\
\bigskip
\begin{small}
$^{\ddagger}$ {\em Department of Mathematics and
Statistics, Concordia University\\ 1455 de Maisonneuve W., Montr\'eal, Qu\'ebec,
Canada H3G 1M8} \\
\end{small}
\end{center}
\begin{center} 
{\bf Abstract}
\end{center}
On the space of matrices with  rational (trigonometric/elliptic) entries there is a well-known Lie-Poisson $r$-matrix  structure. 
The known $r$-matrices are defined on the Riemann sphere (rational), the cylinder (trigonometric), or the torus (elliptic). We extend the formalism to the case of a Riemann surface $\CC$ of  higher genus $g$:  we consider the moduli space of framed vector bundles of rank $n$ and degree $ng$, where the framing consists of a choice of basis of $n$ independent holomorphic sections  that  trivialize the fiber at a given point $\infty\in \CC$.   The cotangent space is known to be identified with the set of Higgs fields,  i.e., one-forms on $\CC$ with values in the endomorphisms of the vector bundle,   with an additional simple pole at $\infty$. The natural symplectic structure on the cotangent bundle of the moduli space induces a Poisson structure on the Higgs fields. Building on Dolgushev’s higher-genus $r$-matrix construction, we identify the kernel with an explicitly computable non-abelian Cauchy kernel and derive the complete dynamical Poisson algebra, including the mixed bracket between the Higgs field and the kernel.   A detailed discussion of the elliptic case including a comparison with the literature, is also provided. 

\vskip 2cm
\setcounter{tocdepth}{4}
\tableofcontents

\section{Introduction}
On the space of square matrices $\Phi(z)$ with entries belonging to the ring of rational functions there exists the well-known Lie-Poisson structure defined by the formula \cite{RS1, RS2, Sem} 
\be
\label{genus0}
\big\{ \Phi_{ab}(z), \Phi_{cd}(w) \big\}  =  \frac{\Phi_{cb}(z)-\Phi_{cb}(w)}{z-w} \delta_{ad}  - \frac{ \Phi_{ad}(z)-\Phi_{ad}(w)}{z-w} \delta_{bc}, 
\ee
or, equivalently, in tensor notation 
\be
\label{genus0ten}
\le\{\m{\Phi}^1(z), \m{\Phi}^2(w) \ri\} = \le[ \m{r}^{12}(z,w)  , \m{\Phi}^1(z) + \m{\Phi}^2(w) \ri],  \ \ \ \ \
\m{r}^{12}(z,w) =\frac { \m{\Pi}^{12}} {z-w}.
\ee
The notation $\m{\Pi}^{12}$ indicates the automorphism of $\C^n\otimes \C^n$ that swaps the terms: $\m{\Pi}^{12}v\otimes u = u \otimes v$. 
The $r$-matrix indicated above is known as the {\it rational} $r$-matrix. Generalizations exist to {\it trigonometric} and {\it elliptic} $r$-matrices, involving eponymous functions in their entries. 

Isospectral flows of $\Phi(z)$ can be obtained by using certain spectral invariants of $\Phi$ as Hamiltonian generators; for example, on the Poisson submanifold of matrices with {\it simple} poles  at $z = x_1,\dots, x_N$ (i.e. the Gaudin model \cite{Gaudin}), one uses $H_j= \res{z=x_j} \frac 1 2 \tr \Phi^2(z) \d z$ as commuting Hamiltonians. Separating Darboux  coordinates for these integrable systems were provided in \cite{AHH}.

Another important example is the rational and elliptic-Calogero-Moser  (eCM) systems; they both have Lax-matrix formulations. 
These isospectral dynamical systems were generalized by Hitchin in \cite{Hitchin} so that the matrix $\Phi$  depends on a point $p$ belonging to an arbitrary smooth Riemann surface $\CC$ (of arbitrary genus). To be more precise, in Hitchin's construction $\Phi$ is a  holomorphic differential with values in ${\rm End} (\scr E)$, where $\scr E$ is a stable  vector bundle over $\CC$ of rank $n$. This is called a {\it Higgs field} and it can be identified with an element of the cotangent space of the moduli space of vector bundles at the point $\scr E$. To recover the rational case one has to consider ``generalized'' Higgs fields (allowing poles) and specialize $\CC$ to $\mathbb P^1$.  In the elliptic case the eCM system can also be  presented as a special case of Hitchin's system \cite{Rubtsov1}.

In the elliptic (and trigonometric) case the $r$-matrix structure \eqref{genus0}  was defined in \cite{BradenSuzuki, Sklyan, Rubtsov1}; for the three cases, rational, trigonometric and elliptic, in fact, there are families of pencils of compatible structures, as shown in \cite{HH}. 

The generalization to higher genus was addressed in \cite{Enriquez1}  and \cite {Felder}: in the former the presentation is in the Schottky uniformization, expressing the $r$--matrix as an infinite series over the Schottky group of the surface (see Sec. 3 ibidem). In the latter, the $r$--matrix is formulated in terms of its properties in the adjoint representation of a semisimple Lie algebra as an existence theorem (sec. 4.1). In both cases the dynamical $r$-matrix equations are stated without direct verification as a consequence of the Jacobi identity.

The question  of an {\it explicit} $r$-matrix structure for Hitchin's systems over a curve of arbitrary genus $g$ is precisely the goal of this paper; moreover we compute the {\it full} Poisson algebra of brackets.  In fact in \cite{HM} the authors conclude ``[...] we can see that the Poisson geometry of rational surfaces suggests quite strongly that there is no nice Poisson extension of the Sklyanin bracket to arbitrary base curves''.  An obstruction of similar nature was observed much earlier in \cite{MikhailovZakharov}. 
{The use of the Tyurin parametrization and a partial description of the $r$--matrix was considered in \cite{Dolg}, which is along the same lines as the aforementioned \cite{Enriquez1, Felder} but with a more intrinsic approach that uses the Riemann surface intrinsically rather than representing it as a Schottky quotient.
We further comment in Remark \ref{fclo} on the added value of the present paper.  
}

Our approach here is rather direct and computational: first of all,  we consider the moduli space, $\mathfrak E$,  of {\it framed} vector bundles of rank $n$ and degree $ng$. The choice of degree is quite natural for several reasons, the principal one being the Riemann--Roch theorem which guarantees the existence, generically, of precisely $n$ independent sections. We provide, following Tyurin \cite{Tyurin}, an explicit parametrization of the moduli in terms of Tyurin data. See also \cite{BertoNortonRuzza}. 

Using the trivialization provided by the $n$ sections, the  framed Higgs fields can be represented concretely as matrices with entries that are meromorphic differentials, see Section \ref{framed}. The reproducing kernel $\Cauchy$ on this space is called the {\it non-abelian Cauchy kernel}, see Def. \ref{defCauchy}. 

Framed Higgs fields are interpreted as elements of $T^\star_{\scr E} \mathfrak E$, and hence the canonical symplectic structure on $T^\star \mathfrak E$ can be written as a Poisson bracket for the Higgs fields. 
This finally leads to  Theorem \ref{main}, providing  a {\it dynamical $r$--matrix structure}  constructed in the most explicit way in terms of the non-abelian Cauchy kernel. It takes the following form, using the common tensor notation:
\bea
\le\{\m{\Phi}^1(p) \m{,}^\otimes\m{\Phi}^2(q)  \ri\}
 &= 
   \le[\m{\Phi}^1(p), \m{r}^{21}(q,p) \ri]
 -\le[\m{\Phi}^2(q), \m{r}^{12}(p,q) \ri]  \nn\\
 \le\{\m{\Phi}^1(w) \m{,}^\otimes\m{\Cauchy}^2(q,r)  \ri\}\
&=
 \m{\Cauchy}^1(w,r)\m{\Cauchy}^2(q,w) \Pi 
+ \m{\Cauchy}^1(w,q)\m{\Cauchy}^1(q,r) \Pi
-\m{\Cauchy}^1(w,r) \m{\Cauchy}^2(q,r) \Pi 
,
\nn\\
 &\m{r}^{12}(p,q):= \m{\Cauchy}^1(p,q) \Pi,\ \ \ \ 
\m{r}^{21}(q,p) :=   \m{\Cauchy}^2(q,p) \Pi,
\label{Poisintro}
 \eea
where $\Pi \in {\rm Aut}(\C^n\otimes \C^n)$ is the permutation operator $\Pi(v\otimes w) = w\otimes v$ and  for $M\in {\rm End} (\C^n)$ we have used $\m{M}^1 = M\otimes \1$, $\m{M}^2 = \1 \otimes M$. The form of the Cauchy kernel matrix $\Cauchy$ is given in Section \ref{seccauchy}.
A detailed  comparison with the well-known results in the case of elliptic curves is provided in Section \ref{example}. 
The proof of Theorem \ref{main} is elementary but rather complicated and is contained in Appendix \ref{proof}. 

\begin{example}
Consider the expression $\H_1(q):= \tr \Phi(q)$; this is a holomorphic differential (all residues at $t_j$, $\infty$ vanish) and a  simple computation from \eqref{Poisintro} yields:
\be
\{\Phi(p), \H_1(q)\} = \le[\Phi(p), \Cauchy(q,p) \ri].
\ee
Note that in genus $0$ this is automatically zero since then $\Cauchy$ is multiple of the identity matrix. 
The periods of $\tr \Phi$ along a Lagrangian basis in homology form a collection of $g$ commuting  and independent Hamiltonian functions and so the corresponding flows are given by the commutator of $\Phi$ with the corresponding matrix of periods of the Cauchy kernel, $M_\gamma(q)= \oint _{p\in \gamma} \Cauchy(p,q)$.

Similarly, consider  $\H_2(q):= \frac 12 \tr \Phi^2(q)$. This is a quadratic differential on $\CC$ with a single double pole at $\infty$, in general, for a framed Higgs field (there are no poles at the Tyurin points). It generates certain Hamiltonians for the Hitchin system.  Again, a simple computation using \eqref{Poisintro} yields
\be
\{\Phi(p), \H(q)\} = \le[\Phi(p), \Phi(q)\Cauchy(q,p) \ri]. 
\ee

\end{example}
\br
\label{fclo}
The original appearance of this $r$-matrix dates back to \cite{Dolg} in the formulation based on Tyurin data, and even earlier in more abstract terms, as already commented upon, in \cite{Enriquez1, Felder}.   Dolgushev defined an $r$-matrix which has the same properties as the one we give in Theorem \ref{main}, but  with two  important practical differences:
\begin{enumerate}
\item the existence of the $r$--matrix is stated in Lemma 1, Lemma 3 ibidem: these are existence  statements for what we call here the non-abelian Cauchy kernel. However, that work  did not provide an explicit formula comparable to \eqref{Cauchyentries} below or the other explicit formulas contained in \cite{BertoNortonRuzza} for arbitrary Tyurin data. The Cauchy kernel also plays other interesting roles as a reproducing kernel which we discuss later in this paper (but this was already introduced in \cite{BertoNortonRuzza}).
\item While it was stated that the $r$--matrix is dynamical, the Poisson brackets with the Higgs field were not evaluated, which is necessary to close the Poisson algebra. This  was in fact recognized in Section 4 ibidem as a missing feature.    
\end{enumerate}
The present paper provides a fully explicit expression for the $r$--matrix and also a complete description of the resulting Poisson algebra. See also App. \ref{appb}.
\er
\paragraph{Acknowledgments.}
The author completed the work during his tenure as Royal Society Wolfson Visiting
Fellow (RSWVF/R2/242024) at the School of Mathematics at the University of  Bristol. The work was supported in part by the Natural Sciences and Engineering Research Council of Canada (NSERC) grant RGPIN-2023-04747.

\section{Framed vector bundles and their moduli}
\label{framed}
Let $\CC$ be a smooth Riemann surface of genus $g$ with complex structure. We consider a generic holomorphic vector bundle $\scr E$ of rank $n$ and degree $ng$. The moduli space of such objects is locally  a smooth complex manifold of  dimension $n^2(g-1)+1$. The corresponding cotangent bundle is canonically a complex-symplectic manifold of twice the dimension: if $q_\alpha$ is a local set of coordinates and $p^\alpha$ the corresponding momenta (coordinates in the cotangent space), the symplectic structure is expressed as 
\be
\Omega:= \d \theta, \ \ \ \theta:= \sum_{\alpha} p^\alpha \d q_\alpha,
\ee
where $\theta$ is the canonical {\it Liouville form}. 
This symplectic structure allows us to construct interesting dynamical integrable systems by choosing a convenient polarization of the  symplectic structure, i.e., a collection of independent hamiltonian functions that commute with one another, as many as the dimension of the underlying manifold. 
The principal example is that of the  Hitchin systems  recalled in the introduction which generalizes the Calogero-Moser system. We review this in more detail in Section \ref{example}.

In this work we are interested in a small refinement of the moduli space \cite{Markmanthesis}: we consider, in fact, {\it framed} vector bundles. By the Riemann-Roch theorem, a vector bundle of rank $n$  and degree $ng$ satisfies
\be
h^0(\scr E) = h^1(\scr E) +  \deg E - n(g-1) = h^1(\scr E) + n.
\ee
Generically $h^1(\scr E)=0$ and hence there are $n$ linearly independent global sections, $\sigma_1,\dots, \sigma_n$. By definition of degree, the section $\sigma_1\wedge \cdots \wedge \sigma_n$ of the determinant bundle has divisor of zeros of degree $ng$; we denote this divisor by $\scr T$ and call the ``Tyurin divisor" \cite{Tyurin}. We will make the genericity assumption that $\scr T$ consists of $ng$ points of multiplicity one. The general case was discussed at length in \cite{BertoNortonRuzza}. 

The ``framing'' consists in fixing a specific basis. A convenient way to realize the framing is to choose a generic point, $
\infty\in \CC \setminus \scr T$ over which the bundle is trivialized and, in this trivialization, require the sections to take a pre-assigned value (we will think of sections as row-vectors). 

For each $t_j\in \scr T$ the determinant $\sigma_1\wedge \dots \wedge \sigma_n$ vanishes to order $1$ thanks to our genericity assumption, which means that there is a vector $\h^{(j)}\in \mathbb P^{n-1} $ such that $\sum \h^{(j)}_a \sigma_a(t_j) = 0$. These are the {\it Tyurin vectors} and  we denote them  collectively by ${\bf H}$ to be thought of as an $n\times ng$ matrix whose columns  are the Tyurin  vectors.
The data of Tyurin points and corresponding vectors provide moduli for  a vector bundle framed over $\infty$, see \cite{Tyurin, BertoNortonRuzza}. Different framings amount to a simultaneous linear transformation of all Tyurin vectors while the Tyurin points are independent of the framing. 
We shall denote by $\frak E = \frak E_{n,ng, \infty}$ the moduli space of such framed vector bundles, which is then of dimension 
\be
\dim \frak E = n^2g. 
\ee
Indeed, we have $ng$ Tyurin points and, for each of those, a point in $\mathbb P^{n-1}$. 

\paragraph{Cotangent space of $\frak E$.}

Since we are framing the bundles at $\infty$, infinitesimal deformations of the bundle must preserve the framing. In duality, a cotangent vector at $\scr E\in \frak E$ can be identified with a {\it framed Higgs field}, namely a section of ${\rm End}(\scr E)\otimes \mathcal K_\CC(\infty)$, where $\mathcal K_\CC$ is the canonical bundle of $\CC$.   This means that a (framed) Higgs field is an endomorphism-valued meromorphic differential with at most a  simple pole at $\infty$. This is an idea that goes back to \cite{Markmanthesis}. 

We will consider a much more concrete representation of a Higgs field in the {\it global} trivialization provided by the framing itself: indeed, if we denote by $\Gamma(p) = [\sigma_1(p), \dots, \sigma_n(p)]^t$ (a column vector of the chosen basis), then a framed Higgs field becomes simply an $n\times n$ matrix consisting of meromorphic differentials 
\bea
\phi: H^0(\scr E) \to  H^0(\scr E\otimes \K_\CC(\infty))\nn\\
\phi(\sigma_a) = \sum_{b=1}^n \Phi_{ab}(p) \sigma_b(p).
\eea
The matrix $\Phi(p) = [\Phi_{ab}(p)]_{a,b=1}^{n}$ consists of meromorphic differentials with the properties:
\begin{enumerate}
\item $\div \Phi\geq -\scr T-\infty$;
\item the residue $\res{p=t_j} \Phi = N^{(j)}$ is a nilpotent matrix of rank at most one of the form  $N^{(j)}_{ab} = \v^{(j)}_a \h^{(j)}_b$, $(N^{(j)})^2 =0$ (i.e. $\v^{(j)}_a \h^{(j)}_a=0$). 
\item the matrix $N^{(j)} \Phi(p)$ is locally analytic at $p=t_j$, by the previous condition  and the evaluation at $p=t_j$ is proportional to $N^{(j)}$. (In the case $N^{(j)}=0$,  the meaningful condition is that ${\h^{(j)}}^t \Phi(t_j) \propto {\h^{(j)}}^t$.)
\end{enumerate}
We recall that the divisor of a matrix  consists of the formal sum of points on $\CC$ counted with the minimum vanishing  order amongst its entries at that point. 
It follows that the local behaviour of a Higgs field at $t_j$ is, in a local coordinate $z = z(p)$,
\bea
\label{Phiexpj}
\Phi(p)  = \le(\frac{N^{(j)}}{z-\zeta_j} + \Phi^{(j),0}  + \Phi^{(j),1}(z-\zeta_j) + \mathcal O(z-\zeta_j)^2\ri) \d z\nn\\
N^{(j)} := \v^{(j)} {\h^{(j)}}^t, \ \ \ (N^{(j)})^2  =0, \ \ N^{(j)} \Phi^{(j),0} = \varkappa_j N^{(j)}.
\eea
Note that the definition of $\varkappa_j$ makes it clear that it represents an element of the canonical bundle of $\CC $ at $p=t_j$.

These conditions first appeared  in \cite{KricheverLax} but we can explain them as the necessary and sufficient conditions for the sections $\{\phi(\sigma_a), \ \ a = 1,\dots,n\}$ to be linearly dependent at $t_j$ with the coefficients of linear dependence given by the same Tyurin vector  $\h^{(j)}$. Since the Tyurin vectors are defined up to multiplicative scalar, so are the vectors $\v^{(j)}$ in \eqref{Phiexpj}: however the matrix $N^{(j)} = \res{p=t_j} \Phi$ is unambiguously defined, and it belongs to the  smooth manifold of rank-one nilpotent matrices, a Poisson manifold of dimension $2n-2$ in ${\frak {gl}}_n$, with respect to the Lie-Poisson structure.

\begin{remark}[Computation of $h^1(\scr E)$.] \rm 
\label{remTyurin}
It was clarified in \cite{BertoNortonRuzza} how to concretely compute $h^1(\scr E)$; under our genericity assumptions (see ibidem for the general formulation) the dimension is given by 
\be
h^{1}(\scr E)  = {\rm corank} \, \mathbb T
\ee
where $\mathbb T$ is a matrix of size $ng\times ng$, the {\it Brill-Noether-Tyurin matrix}, given by the expression:
\begin{equation}
\label{TyurinT}
\mathbb T:= \le[\omega_1(\scr T)\mathbf H^t|  \omega_2 (\scr T) \mathbf H^t 
| \cdots |  \omega_g (\scr T) \mathbf H^t\ri]
\end{equation}
where $\{\omega_1,\dots, \omega_g\}$ is any basis of holomorphic differentials and 
\begin{equation}
\mathbf H = [\mathbf h^{(1)}| \cdots | \mathbf h^{(ng)}]\in \ {\rm Mat}(n\times ng)\qquad
\omega_\ell(\scr T):= {\rm diag} \Big(\omega_{\ell}(t_1), \dots, \omega_{\ell}(t_{ng})\Big)\in \Mat(ng,ng).
\end{equation}
Note that, as it should, the corank of $\mathbb T$ is independent of the framing and of the choice of basis of holomorphic differentials. 
The locus in the moduli space $\frak E$ where $h^1(\scr E)\geq 1$ is called the {\it non-abelian Theta divisor}; in the case of line bundles $n=1$ this is precisely the Theta divisor in the Jacobian. See, for example, \cite{Fay92} 
\er

The moduli space of unframed bundles is obtained as a quotient under the left ${\rm GL}_n$ action of change of basis of sections. The corresponding cotangent space can be identified with the subspace of framed Higgs fields that are holomorphic at $\infty$. In general, the framed Higgs fields are simply matrices of third-kind differentials with poles at $\scr T$ and $\infty$; the residue at $\infty$ is thus
\be
\res{p=\infty}\Phi = - \sum \res{p=t_j} \Phi = -\sum N^{(j)}.
\ee
The condition for $\Phi$ to be a holomorphic Higgs field  and hence be identifiable with a cotangent vector of the unframed moduli space, is that the residue vanishes: this  imposes the  $n^2-1$ constraints  $\sum N^{(j)}=0$ (each $N^{(j)}$ is already traceless) so that the dimension of these holomorphic Higgs fields is the correct one $n^2 g - n^2+1 = n^2(g-1)+1$.  
\subsection{The reproducing kernel on the space of framed Higgs fields}
\label{seccauchy}
The main tool for extending the r-matrix structure is the {\it non-abelian Cauchy kernel}, introduced in \cite{BertoNortonRuzza}.  A similar notion appears in Section 3 of \cite{HurBis}.
For $h^1(\scr E) =0$, that is,  away from the non-abelian Theta divisor, we can describe the Cauchy kernel as follows.  For $q\in \CC$ we have $ h^0(\scr E^\vee\otimes  \K_\C(q+\infty)) = n$: the $n$ sections spanning  $H^0(\scr E^\vee\otimes  \K_\C(q+\infty)) $ can be arranged into a matrix $\mathbb K(p;q)\in H^0(\scr E^\vee \otimes \scr E\otimes \K_\CC(q+\infty))$; normalizing it to have residue representing the identity on the fiber at $q$ yields the non-abelian Cauchy kernel. 
In keeping with our desire for explicit formulas, let us trivialize the above kernel using the basis of sections $\Gamma$ as above. Then $\mathbb K(p;q) = \Gamma(p)^{-1} \Cauchy(p,q) \Gamma(q)$, where $\Cauchy(p,q)$ is just a matrix of third-kind differentials with respect to $p$ and poles at $p=q,\infty$ of special form; see \cite{BertoNortonRuzza}.

In concrete terms, under our genericity assumption,  the non-abelian Cauchy kernel $\Cauchy(p,q)$ boils down to the definition below.
\bd[Cauchy kernel]
\label{defCauchy}
The Cauchy kernel $\Cauchy(p,q)$  is characterized by the properties
\begin{enumerate}
\item It is a third-kind meromorphic differential in the variable $p$ and a meromorphic function in the variable $q$; 
its only poles as a differential in $p$ are at $p =q$ and $p=\infty$ and 
\begin{equation}
\res{p=q} \Cauchy(p,q) =\1 = -\res{p=\infty} \Cauchy(p,q).
\end{equation}
\item 
For all $j=1,\dots, ng$ we have 
\be
\label{hjCauchy}
\h^{(j)} \Cauchy(t_j,q) \equiv 0, \ \ \ \forall q\in \mathcal C.
\ee
\item Let $z_j$ be a local coordinate near $t_j$ so that $z_j(t_j) = \zeta_j$; then, as a function of $q$ the kernel $\Cauchy$ has a simple pole and singular part of rank one with the same row space spanned by the Tyurin vector, as follows;
\be
\label{singpartCauchy}
\Cauchy(p,q) = \frac{\f^{(j)}(p){ \h^{(j)}}^t}{z_j(q)-\zeta_j} + \mathcal O(1) .
\ee
The (column) vector $\f^{(j)}(p)$ consists of one-forms.
\end{enumerate}
\ed
We can write the entries of $\Cauchy$ as follows, with the same notation as Remark \ref{remTyurin}: 
\bea
\label{Cauchyentries}
\Cauchy_{ab}(p,q)=\frac{ \det \le[\begin{array}{c|c|c|c}
\begin{array}{c}
\\
\omega_1(\scr T) \mathbf H^t
\\
\phantom{j}\end{array}  & \cdots &\begin{array}{c}
\\
\omega_g(\scr T) \mathbf H^t
\\
\phantom{j}
\end{array}  & 
\omega_{q\, \infty} (\scr T) \mathbf H^t \mathbf e_b\\
\hline
 \omega_1(p){\mathbf e}_a^t & \cdots & \omega_g(p) \mathbf e_a^t & \omega_{q,\infty}(p)\delta_{ab}
\end{array}\ri]
}
{\det \le[\begin{array}{c|c|c}
\begin{array}{c}
\\
\omega_1(\scr T) \mathbf H^t
\\
\phantom{j}\end{array}  & \cdots &\begin{array}{c}
\\
\omega_g(\scr T) \mathbf H^t
\\
\phantom{j}
\end{array}  
\end{array}\ri]}
\eea
Here $\omega_1,\dots,\omega_g$ are holomorphic differentials normalized  along   $g$ mutually nonintersecting, homologically independent cycles, $\oint_{a_j} \omega_k = \delta_{jk}$,  and $\omega_{q,\infty}(p)$ is the unique third-kind differential with poles at $p=q,\infty$ and residues $1,-1$, respectively, and with vanishing periods along the same cycles. 
Note that this expression makes it clear that the Cauchy kernel is single--valued as a meromorphic function of $q$; analytic continuation along the $b_\ell$ cycle adds to the last column one of the previous columns. In particular it is independent of  the choice of cycles used to normalize the differentials. 
\br
The formula \eqref{Cauchyentries} may seem to be  computationally transcendental because it requires the computation of the normalized differentials and hence the matrix of periods along a symplectic basis in homology. This is actually not the case, and  we provide an alternative choice of basis which is purely algebraic (i.e. does not involve computation of periods)  in App. \ref{appb}. 
\er

{\bf Notational convention:} since many computations rely on the use of a local coordinate $z_j(x)$ for $x$ near a Tyurin point $t_j$, we shorten the notation by simply appending an index, so that $x_j$ means $z_j(x)$, etc.

\bc
The vectors $\f^{(j)} (p)$ of one-forms  solve the following interpolation problem:
\begin{enumerate}
\item ${\h^{(k)}}^t \f^{(j)} (t_k) = 0$ for $k\neq j$;
\item $\frac {{\h^{(j)}} ^t \f^{(j)}(p)}{\d p_j}
\bigg|_{p=t_j} = 1$, where $p_j=z_j(p) $ is the same local parameter used in the definition of the singular part in \eqref{singpartCauchy} .
\end{enumerate}
\ec
{\bf Proof.} This can be seen directly from the determinantal formula \eqref{Cauchyentries} or as a consequence of the definition. While the first property is immediate from the definition, the second requires a bit more careful analysis. If $p,q$ are both in the coordinate patch covered by $z_j$, then we can write $p_j= z_j(p), q_j = z_j(q)$ and 
\be
\Cauchy(p,q) = \frac{ \f^{(j)}(p) {\h^{(j)}}^t}{q_j-\zeta_j} + \frac { \d p_j}{p_j-q_j} + H(p_j,q_j)\d p_j
\ee
where $H(p_j,q_j)$ is a jointly analytic matrix-valued function in the same neighbourhood.
Multiplying on the left by ${\h^{(j)}}^t$ and taking the limit $p\to t_j$ we see that the second property must also hold, for otherwise there would be a singular part. 
\QED
\def\QQ{\mathscr Q}

The Cauchy kernel is the  reproducing kernel in the space of Higgs fields in the following sense:  
\be
\label{reproducing}
\sum_{j=1}^{ng} \res{x=t_j} \Cauchy(p,x)\Phi(x)  = \Phi(p),
\ee
which is a simple application of Cauchy's residue theorem observing that $x=p$ is  the only remaining pole of the differential being taken the residue of. 

Moreover, the Cauchy kernel can be used to {\it construct} a Higgs field with given parameters as explained in the next lemma.
\begin{shaded}\bl
\label{lemmarecovery}
The Higgs field $\Phi(p)$ can be written as follows 
\be
\label{Phirep}
\Phi(p) = \sum_{j=1}^{ng}\res{q=t_j}  \Cauchy(p,q) \le(\frac {N^{(j)}}{q_j-\zeta_j} + \varkappa_j \1 \ri)\d q_j
\ee
(see the notational convention above for the meaning of $q_j$)
where the parameters $\varkappa_j$  and the rank-one nilpotent matrices  $N^{(j)}$ are arbitrary.
\el
\end{shaded}
\noindent{\bf Proof.}
One verifies directly that the two sides have the same residues at $t_j$, and moreover $N^{(j)}\Phi(t_j) = \varkappa_j N^{(j)}\d \zeta_j$. To verify the last assertion one must take into account that the limit $p\to t_j$ does not commute with taking the residue, because the Cauchy kernel has a pole on the diagonal and rather one should compute it  as follows; in the $z_j$ coordinate neighbourhood of $t_j$ we have
\bea
\res{q=t_j}  \Cauchy(p,q) \le(\frac {N^{(j)}}{q_j-\zeta_j} + \varkappa_j \1 \ri)\d  q_j = \le( \frac {N^{(j)}}{p_j -\zeta_j} +\varkappa_j \1 \ri) \d p_j + \frac 1{2i\pi} \oint \Cauchy(p,q) \le(\frac {N^{(j)}}{q_j-\zeta_j} + \varkappa_j \1 \ri)\d q_j
\eea
where the contour of integration is chosen so that  $p$ lies inside the disk centered at $t_j$. Then multiplying by $N^{(j)}$, the singular term vanishes because $(N^{(j)})^2=0$, while the last term vanishes at $p=t_j$ because of the property \eqref{hjCauchy} of $\Cauchy(p,q)$ and the fact that the row-span of the matrix $N^{(j)}$ is  proportional to $\h^t_j$. 
 Since these equations characterize  $\Phi$,  the lemma follows.
 \QED

\paragraph{Mittag-Leffler problems.}
The Cauchy kernel $\Cauchy$ solves two problems of Mittag-Leffler type: suppose we are looking for a meromorphic section, $\sigma$, of $\scr E$ with a prescribed principal part at a point $r_0$:
$$
\sigma(q)= \sum_{\ell=0}^L \sigma_{\ell} \frac 1{z^\ell} + \mathcal O(|z|) =\sigma_{sing}(z) + \mathcal O(1),
$$
here $z = z(q)$ is  a local coordinate centered at $r_0$ ($z(r_0)=0$). 
Then the Cauchy kernel allows us to turn this local information into a global one by means of 
\be
\sigma(q) = \res{p=r_0} \sigma_{sing}(z(p)) \Cauchy (p,q),
\ee
providing the unique section having the prescribed singular expansion and  vanishing at $\infty$. 
On the dual side, if we are looking for a meromorphic section $\phi$  of $\scr E^\vee \otimes \K_\CC$
then a similar task is accomplished with the aid of the formula
$$
\phi(p) = \res{q=r_0} \Cauchy(p, q) \phi_{sing}(z(q)) \d z(q). 
$$

\section{The higher genus  $r$-matrix structure}
It was shown by Krichever \cite{KricheverLax} that the symplectic structure on $T^\star \mathfrak E$ induces the following Poisson structure on the coordinates $N^{(j)}, \zeta_j, \varkappa_j$:
\bea\label{Npo}
\{N^{(j)}_{ab}, N^{(k)}_{cd} \} &= \delta_{jk} \le(N^{(j)}_{ad} \delta_{cb} - N^{(j)}_{cb} \delta_{ad}\ri),\ \ \ 
{\{\varkappa_j, \zeta_j\} =  -1},
\eea
with all other Poisson bracket vanishing.
We want to express the Poisson bracket in closed form for the Higgs field $\Phi$, similarly to the case of the $r$-matrix structure on $\mathbb P^1$.
We recall that if $\Phi(z)$ is a matrix-valued rational function, the Lie-Poisson bracket is defined by \eqref{genus0}. 
We promote $\Phi$ to a matrix-valued differential to keep track of the tensorial weights under change of coordinate. If $\Cauchy_{ab}(z,w)=\frac{\d z}{z-w} \delta_{ab}$ denotes the Cauchy kernel in genus zero, then \eqref{genus0}  can be rewritten
\be
\label{wrongansatz}
\big\{ \Phi_{ab}(z), \Phi_{cd}(w) \big\} = \Phi_{ad}(z)\Cauchy_{cb}(w,z) -\Phi_{ad}(w)\Cauchy_{cb}(z,w)
-\Phi_{cb}(z) \Cauchy_{ad}(w,z) + \Phi_{cb}(w) \Cauchy_{ad}(z,w).
\ee
Note the order of arguments of $\Cauchy$ which is dictated by the fact that the left and right sides should be bi-differentials.
It is tempting to use the exact same expression \eqref{wrongansatz} but with the new understanding of $\Phi$ as a framed Higgs field and $\Cauchy$ as the non-abelian Cauchy kernel.
 However this is certainly not correct as we explain presently.  Let us see what structure the pole at $p=t_j$ should have; the bracket  must start with a double pole with leading singular part $(\v_j \h_j^t)_{ab}$ because $t_j$ is coupled to the eigenvalue $\varkappa _j$. The first term has this structure in the $(a,b)$ indices, but there is also the third term with a double pole and the leading coefficient is $(\f^{(j)}(q) {\h^{(j)}}^t)_{ad}$ which is incorrect. 
Instead, we will prove
\bt
\label{main}
The canonical symplectic structure on the cotangent bundle of the moduli space of framed vector bundles of rank $n$ and degree $ng$ over $\CC$ induces the following Poisson brackets:
\bea
\label{PhiPhi}
\Big\{\Phi_{ab}(p) ,\Phi_{cd}(q) \Big\}= 
\PhiPhi abpcdq
\eea

\bea
\label{PhiK}
\Big\{\Phi_{ab}(w) ,\Cauchy_{cd}(q,r) \Big\}= \PhiK abwcdqr ,
\eea
where $\Cauchy(p,q)$ is the non-abelian Cauchy kernel of Def. \ref{defCauchy}  and with entries \eqref{Cauchyentries}.
In the St. Petersburg notation, this reads 
\bea
\label{PhiPhiten}
\le\{\m{\Phi}^1(p) \m{,}^\otimes\m{\Phi}^2(q)  \ri\}
 &= 
   \le[\m{\Phi}^1(p), \m{\Cauchy}^2(q,p) \Pi \ri]
 -\le[\m{\Phi}^2(q), \m{\Cauchy}^1(p,q) \Pi \ri]  \\
\label{PhiKten}
 \le\{\m{\Phi}^1(w) \m{,}^\otimes\m{\Cauchy}^2(q,r)  \ri\}\
&=
 \m{\Cauchy}^1(w,r)\m{\Cauchy}^2(q,w) \Pi 
+ \m{\Cauchy}^1(w,q)\m{\Cauchy}^1(q,r) \Pi
-\m{\Cauchy}^1(w,r)\m{\Cauchy}^2(q,r) \Pi ,
\eea
where $\Pi \in {\rm Aut}(\C^n\otimes \C^n)$ is the permutation operator $\Pi(v\otimes w) = w\otimes v$ and  for $M\in {\rm End} (\C^n)$ we have used $\m{M}^1 = M\otimes \1$, $\m{M}^2 = \1 \otimes M$.

 \et
We observe that 
in genus $0$  the formula \eqref{PhiPhi} gives the standard result and \eqref{PhiK} is zero on account of the trivial algebraic identity
$$
\frac 1{w-r} \frac 1{q-r} - \frac 1{w-r} \frac 1 {q-w} - \frac 1{w-q} \frac 1 {q-r} \equiv 0.
$$
Moreover the expressions do  have no singularities for $p=q,q=r,r=p$, as it should be.

The proof of Theorem \ref {main} is contained in App. \ref{proof} and   proceeds by computing the individual Poisson brackets. We then prove separately  Theorems \ref{thmKPhi} and \ref{thmPhiPhi}, which together are equivalent to Theorem \ref{main}.

The Poisson brackets \eqref{PhiPhiten}, \eqref{PhiKten} necessarily satisfy the Jacobi identity because they will be derived directly from the Poisson brackets \eqref{Npo}. In particular, the corresponding dynamical Yang-Baxter-type relation holds:
\bea
\le[ \m{\Phi}^1(p), 
\le[\m{r}^{12}(p,q),\m{r}^{13}(p,s) \ri] + 
\le[\m{r}^{12}(p,q),\m{r}^{23}(q,s) \ri] + 
\le[\m{r}^{32}(s,q),\m{r}^{13}(p,s) \ri] + 
\le\{\m{\Phi}^2(q), \m{r}^{13}(p,s) \ri\}
-\le\{\m{\Phi}^2(s), \m{r}^{12}(p,q) \ri\}\ri] \nn\\
+ 
\text{cyc. perm.} =0 .\nn
\eea
It is remarkable, however,  that a direct verification appears significantly complicated.
In view of this difficulty it is therefore not surprising that the computation of the Poisson brackets $\{\Phi, \Cauchy\}$ was not completed in \cite{Dolg}. The proof in Section \ref{proof} is not particularly elegant but it is straightforward, relying heavily on the properties of the non-abelian Cauchy kernel. 

Although this fact  is quite well known, it {\it follows} from \eqref{PhiPhiten} that the eigenvalues of $\Phi$, i.e., the moduli of the spectral curve $\wh \CC= \{\det (\lambda\1 - \Phi)=0\}$ embedded in $\K_\CC$ Poisson commute, 
$$
\{\l(\wh p), \l (\wh q)\} =0, \ \ \ \wh p, \wh q\in \wh \CC. 
$$
 Indeed we can rewrite \eqref{PhiPhiten} as 
 \bea
 \le\{\m{\Phi}^1(p) \m{,}^\otimes\m{\Phi}^2(q)  \ri\}
 &=  \le[\m{\Phi}^1(p), \m{r}^{21}(q,p) \ri] - \le[\m{\Phi}^2(q), \m{r}^{12}(p,q) \ri] \\
 &\m{r}^{12}(p,q):= \m{\Cauchy}^1(p,q) \Pi,\ \ \ \ 
\m{r}^{21}(q,p) :=   \m{\Cauchy}^2(q,p) \Pi,
 \eea
 and then a standard computation (see e.g. the Proposition on pag. 14 in  \cite{BabelonBook}) implies the involutivity of the eigenvalues.
 The genus $\wh g$ of $\wh \CC$ is, for a generic framed Higgs field $\Phi$, exactly equals  the dimension of the moduli space $\frak E$ of framed vector bundles. 

\paragraph{Change of framing.}
Since our framing consists in fixing the evaluation of sections at an arbitrary point $p=
\infty$, the change of framing amounts to a left ${\rm GL}_n$ action on the matrix of sections. The Higgs fields  then transform under the adjoint action of ${\rm GL}_n$. It is immediate from \eqref{Npo} that the Hamiltonian generator of such an action is the matrix 
\be
\mathbb J = \sum_{j=1}^{ng} N^{(j)} =-\res{p=\infty} \Phi(p). \label{genframechange}
\ee
Namely  the flow ${\rm e}^{tF}\in {\rm GL}_n(\C)$ of frame changes is generated by the Hamiltonian $H:= \tr (F\mathbb J)$.

\section{Example: the elliptic case revisited}
\label{example}
The r-matrix structure on an elliptic curve was derived in \cite{BradenSuzuki, Sklyan, Rubtsov1}. To be as concrete as possible we realize the elliptic curve as $\mathcal E_\tau = \C/\Z+\tau \Z$, using the symbols $p, q,\dots$ as global coordinates rather than abstract points.

In genus $g=1$ the moduli space of unframed bundles of rank $n$ and degree $ng = n$ is rigid, but  the framed moduli space has  dimension  $n^2$ arising entirely from the framing. For a framed bundle $\scr E$ we have $n$ Tyurin points $t_1,\dots t_n$ (distinct as per our genericity assumption) and correspondingly $n$ Tyurin vectors. 
The condition that  $h^0(\scr E)=n$ translates simply into the linear independence of the Tyurin vectors and so, up to a reframing of the bundle, we can assume without loss of generality that the Tyurin vectors are $\h^{(j)}\propto  \e_j$, the unit vectors.  We use the ordinary $\theta_1(z) = \theta_1(z|\tau)$ function with the properties
\be
\theta_1(z+1) = -\theta_1(z); \ \ \theta_1(z+\tau) = -{\rm e}^{i\pi\tau - 2i\pi z}\theta_1(z).
\ee
We also use the notation $\zeta(p):= \frac {\d}{\d p}\ln  \theta_1(p)$. 
Let $T = {\rm diag} (t_1,\dots t_n)$ and let us understand that whenever we have a scalar function $f(z)$ then $f(T)= {\rm diag} (f(t_1), \dots, f(t_n))$. We also use $\mathbb H $ for the matrix whose rows are the Tyurin vectors. As per the above discussion we can assume also that it is an arbitrary diagonal invertible matrix (which we could decide to fix as the identity thus ``gauge-fixing'' the construction), but we keep writing it explicitly  to track down the Tyurin vectors' dependence. 
A direct computation shows that the non-abelian Cauchy kernel is provided by  the expression
\be
\Cauchy(p,q) = \Big(\big(\zeta(p-q)- \zeta(q) \big)\1 - \mathbb H^{-1} \le(\zeta(T-q) - \zeta(T)\Big) \mathbb H\ri)\d p.
\ee
Indeed one can directly verify that this expression satisfies the properties for the Cauchy kernel in Def.  \ref{defCauchy}. 
A Higgs field is then a  matrix that ultimately can  be written in  the form
\bea
\Phi(p) &=\mathbb H^{-1}\le( K + \frac{\theta_1(p-T)}{{\theta_1(T)}} 
F(p) \le(\frac{\theta_1(p-T)}{\theta_1(T)} \ri)^{-1} \ri) \mathbb H,\nn\\
&K:= {\rm diag} (\varkappa_1,\dots,\varkappa_n),\ \ \ 
F_{ab}(p) =  C_{ab} \frac{\theta_1(p+t_a-t_b)\theta_1'(0)}{\theta_1(t_b-t_a) \theta_1(p)}, \ \ a\neq b, \ \ \ 
F_{aa}=0.
\label{genPhi1}
\eea
The matrices $N^{(j)}$ are read off as follows
\be
N^{(j)} = \res{p=t_j} \Phi(p) = \mathbb H^{-1}\c_{j} {\h^{(j)}}^t ,
 \ \ \ \tr(N^{(j)}) =0,
\ee 
where $\c_j$ is the $j$-th column of the diagonal--free matrix $C$. Thus, to match with the notation of the previous  sections, $\v^{(j)} = \mathbb H^{-1}\c_{j}$.
Note that 
\be
\res{p=0}\Phi = \sum_a N^{(a)}  =  \mathbb H^{-1} C \mathbb H=: \mathbb J,
\ee
where $C$ is an arbitrary diagonal free matrix. The matrix $\mathbb J$ is the Hamiltonian generator of change of framing as explained in \eqref{genframechange}.

\paragraph{Gauge fixing and comparison with the literature.}
In the gauge where $\mathbb H=\1$, the entries read
\be
\Phi_{ab}(p) = \varkappa_a \delta_{ab} +(1-\delta_{ab}) C_{ab}  \frac {\theta_1(p-t_a)\theta_1(p+t_a-t_b)
\theta_1'(0)\theta_1(t_b)}{\theta_1(t_b-t_a)\theta_1(p-t_b) \theta_1(p)\theta_1(t_a)}.
\ee
Consequently,  the $C_{ab}$ Poisson-commute with the $\varkappa_a, t_a$'s.
In the literature \cite{Rubtsov1} the presentation is different, because instead of a bundle of degree $n$ they use a direct sum of $n$ bundles of degree $0$. It is simple to move from one picture to the other. Apart from the use of multiplicative notation for the elliptic curve in \cite{Rubtsov1}, the Higgs fields in the presentation of loc. cit. are (in our notation  and in the same gauge $\mathbb H=\1$) 
\be
\wt \Phi_{ab} (p) = \varkappa_a \delta_{ab} +(1-\delta_{ab}) C_{ab}  \frac {\theta_1(p+t_a-t_b)}{ \theta_1(p)}.
\ee
Thus the difference between \cite{Rubtsov1} and \eqref{genPhi1} amounts to the absence of the conjugation by $\theta_1(p-T)/\theta_1(T)$ and now the Higgs field is not an elliptic meromorphic function but satisfies  the periodicity 
\be
\wt \Phi(p+1) = \wt \Phi(p), \ \ \wt \Phi(p+\tau) ={\rm e}^{-2i\pi T}  \wt \Phi(p) {\rm e}^{2i\pi T}.
\ee
Up to these minor aesthetical differences, our  $r$--matrix structure in Theorem  \ref{main} then  coincides with that of \cite{Rubtsov1}. It should be  noted  that,  harmless as the conjugation may be, it changes the form of the $r$-matrix so that, in loc. cit.,  the structure takes a slightly more complicated form. 
\subsection {Elliptic Calogero-Moser leaf}
It is also appropriate here to comment on the now venerable construction of Krichever \cite{Krichever1980} of the Lax matrix for the elliptic Calogero-Moser (eCM) systems\footnote{Krichever used the Weierstrass functions for the construction while here we use the Jacobi theta functions, but it is a minor difference; we are here aligned with the presentation of Takasaki \cite{Takasaki1}.}.
The  eCM system is the dynamical system with Hamiltonian 
\be
H = \frac 1 2\sum_{j} \varkappa_j^2 + \frac {\gamma ^2}2\sum_{j\neq i} \wp(t_i-t_j),\ \ \ \{\varkappa_i, t_j\} = \delta_{ij}. 
\ee
Krichever then showed that this can be written as a Lax formulation, which in our notation is as follows.
Consider the Hamiltonian 
\be
\label{Hitchinham}
H:=\frac 1 2  \int_{\mathcal A} \frac{\tr (\Phi^2(p))}{\d p},
\ee
where the $\mathcal A$-cycle is represented as a horizontal segment $[c, c+1]$.
One sees that, since the matrix $F$ is diagonal--free,  
\be
\tr(\Phi^2(p)) =\le( \tr K^2 + \tr F(p)^2 \ri)\d p^2= \le(\sum \varkappa_a^2 - \sum_{a\neq b} C_{ab}C_{ba} \frac{
\theta_1(p+t_a-t_b) \theta_1(p+t_b-t_a)\theta_1'(0)^2}{\theta_1(p)^2 \theta_1(t_a-t_b)^2}\ri) \d p^2. 
\ee
To simplify the expression in the bracket, we observe that it is a symmetric and elliptic  function of the variables  $t_a$. Consider its properties as a function of $t_1$: it has double poles at $t_a,  \ a\geq 2,$ with leading coefficient $C_{1a}C_{a1}$. At $t_2$ (and $t_3,\dots$), it has no residue since the expression is patently even in the exchange $t_1\leftrightarrow t_2$; thus, with $\wp (p) = \frac {\d^2}{ \d p^2} \ln \theta_1(p)$, as a function of $t_1$ it behaves like $\sum_{a>1} C_{1a}C_{a1} \wp(p-t_a) +$ a constant that may depend in principle on $t_2,\dots. $. However since the result must be symmetric, this constant cannot depend on any of the $t_a$ (for otherwise it would also depend on  $t_1$). Similarly, with respect to $p$ it is an even function with double pole at $p=0$ and leading coefficient $\sum_{a\neq b} C_{ab}C_{ba}$. Thus 
\be
\frac{\tr(\Phi^2(p))}{\d p} = \le(\sum \varkappa_a^2 - \sum_{a\neq b} C_{ab}C_{ba} \wp(t_a-t_b) + \sum_{a\neq b} C_{ab}C_{ba} \wp(p) \ri)\d p
\ee
and then the Hitchin's hamiltonian \eqref{Hitchinham} reads 
\be
H= \frac 1 2 \sum\varkappa_a^2 - \frac 1 2 \sum_{a\neq b} C_{ab} C_{ba} \wp(t_a-t_b),
\ee
up to an additive term that is independent of $\varkappa_a, t_a$'s. 
With the Poisson structure of Theorem \ref{main} the flow it generates is just one of the Hitchin flows. The connection with the eCM is obtained if one chooses a particular leaf corresponding to the choice of $C_{ab} = i\gamma $, for all $a,b=1,\dots n$. Namely:
\be
\label{ellcalmat}
L_{ab}(p) = \varkappa_a \delta_{ab}  + i\gamma (1-\delta_{ab})  \frac {\theta_1(p-t_a)\theta_1(p+t_a-t_b)}{\theta_1(p-t_b) \theta_1(p)}.
\ee
Note that this is not immediately Krichever's Lax matrix (beside the use of  Weierstrass instead of Jacobi theta functions). To obtain a form closer to Krichever's,  one needs to conjugate  this $L$ by the diagonal matrix
\be
G = \theta_1(p-T) {\rm e}^{-\zeta(p)T},
\ee
so that 
\be
\wt L(p) = G^{-1}L(p)G, \ \ \ 
\wt L_{ab}(p) = \varkappa_a \delta_{ab}  + i \gamma (1-\delta_{ab}) \frac{\theta_1(p + t_a-t_b)}{\theta_1(p) \theta_1(t_b-t_a)} {\rm e}^{\zeta(p) (t_a-t_b)}.
\ee
Note that this matrix is still elliptic, but with an essential singularity at $p=0 +\Z+\tau\Z$.

The choice of $C_{ab} = i\gamma$ can be explained  more generally by  recalling the formulation of rational Calogero-Moser system and the Kazhdan-Kostant-Sternberg reduction scheme \cite{Kazhdan}. Indeed in that case one starts with the symplectic manifold of pairs of Hermitian matrices  $\mathbb P,\mathbb Q$ with the symplectic structure $\{\mathbb P_{ab}, \mathbb Q_{cd} \} = \delta_{ad} \delta_{cb}$. This is simply the canonical symplectic structure on the cotangent bundle of the vector space of Hermitian matrices.  Then consider the trivial Hamiltonian $H = \frac 1 2\tr \mathbb P^2$ generating the linear flow, $\mathbb Q(t) = t\mathbb P +\mathbb Q$.  The angular momentum $\mathbb J:= [\mathbb P, \mathbb Q]$ is conserved. The Calogero leaf  is obtained by fixing the angular  momentum \cite{Kazhdan}
$$
[\mathbb P,\mathbb Q]_{ab}  = -i\gamma \1 + \text{ rank-one perturbation}
$$
On this coadjoint leaf the eigenvalues of $\mathbb Q(t) = t\mathbb P +\mathbb Q$ satisfy the Calogero dynamics. 

To make the parallel with  the elliptic case explicit, let us reintroduce the arbitrary Tyurin vectors so that the Higgs field is as in \eqref{genPhi1}. As explained in the discussion surrounding  \eqref{genframechange},  the generator of the $Ad_{{\rm GL}_n}$ action is precisely $\mathbb J = \res{p=0}\Phi(p)$. The Lax matrix for the eCM \eqref{ellcalmat} then satisfies exactly the same condition: 
\be
\res{p=0}\Phi (p)\d p = \mathbb J = \mathbb H^{-1} C\mathbb H = i \gamma\mathbb H^{-1} \le(  {\bf u} {\bf u}^t - \1\ri) \mathbb H, \ \ \  {\bf u}^t =[1,1,\dots 1]. \label{Kazhdan}
\ee

\section{Concluding remarks}
The non-abelian Cauchy kernel introduced in \cite{BertoNortonRuzza} clearly  plays a pivotal role in this $r$--matrix structure.  The next natural step is to investigate isomonodromic equations; these were also studied  by Krichever \cite{Kricheveriso}. Our goal would be to arrive at a proper definition of {\it isomonodromic tau function} for such higher-genus monodromy-preserving equations.  
In this direction the groundwork was laid in \cite{BertoNortonRuzza} and the gist is that, using the Cauchy kernel $\Cauchy(p,q)$ one can define unambiguously an affine connection on sections of $\scr E\otimes \sqrt{\K_\CC}$ by taking the regular part of $\Cauchy(p,q)$ along the diagonal $p=q$.  We call this  the {\it Cauchy connection} $\nabla_0$. Connections on $\scr E\otimes \sqrt{\K_\CC}$  form an affine space modelled on the space of  Higgs fields, and hence provide a way to convert a Higgs field to a connection $\nabla = \nabla_0 + \Phi$ and thus define a {\it monodromy} map from the cotangent space of framed bundles,  $T^\star \frak E$, into the ${\rm GL}_n$ character variety. 

Another interesting avenue concerns the relation to the Calogero-Moser system. It was shown in  \cite{KricSam} that the spectral curve of the eCM system \eqref{ellcalmat} cuts a particular locus on the moduli space, in that it admits a meromorphic second-kind differential with a unique double pole and {\it integer periods}. It is quite tempting to define the higher genus Calogero-Moser leaf for framed bundles by the condition that $\res{p=\infty}\Phi = $ rank-one perturbation of the identity matrix, in parallel with \eqref{Kazhdan}.  Then the natural question would be to investigate, in the same spirit as \cite{KricSam}, the properties of the spectral curve for such generalized Calogero-Moser Higgs fields. 
We plan to investigate these issues in forthcoming publications.
\appendix
\renewcommand{\theequation}{\Alph{section}.\arabic{equation}}
 
\section {Proof of Theorem \ref{main}}
\label{proof}
We need to start with the elementary Poisson brackets  which were computed by Krichever in \cite{KricheverLax}.
\bea
\Big\{N^{(j)}_{ab} , N^{(\ell)}_{cd}\Big\}& =  \delta_{j\ell} \Big(\NN jabcd \Big),\qquad 
{\Big\{\varkappa_{j} , \zeta_\ell \Big\} = -\delta_{j\ell}}\label{Poiss1}
\eea
and all other brackets vanish.
\begin{remark}[Reminder of notational convention]
In a neighbourhood of each Tyurin point $t_j\in \CC$ we choose, arbitrarily but once and for all, local coordinates $z_j$, and set $\zeta_j = z_j(t_j)$. Moreover, if $p$ denotes a  point in the same neighbourhood, we shall write in abridged form $p_j$ instead of  $z_j(p)$. The final proof is independent of the choice of coordinates and thus this is a purely instrumental choice. 
\end{remark}
We  prove the following Proposition:
\bp
\label{PropPhiPhi}
The Poisson brackets \eqref{Poiss1} imply
\bea
 \Big\{\Phi_{ab} (p),N^{(j)}_{cd} \Big\} =& \res{q=t_j} \le(\Cauchy(p,q)\Phi(q)\ri)_{ad} \delta_{cb}
 - \res{q=t_j} \Big(\Cauchy_{ad}(p,q)\Phi_{cb}(q)\Big)
 \nn\\
{ \Big\{\Phi(p),\zeta_j\Big\} =}& 
{-\res{q=t_j} \Cauchy(p,q) \d q_j}
 \nn\\
{ \Big\{\Phi(p), \varkappa _j\Big\} =}&
  {-\res{q=t_j} \Cauchy(p,q) \d_q \le(\frac{\Phi(q)}{\d q_j}\ri).} 
\eea
\ep
\noindent{\bf Proof.}
{\bf Poisson brackets with $N^{(j)}$.}
Consider the expressions
$\Big\{\Phi_{ab}(p),N^{(j)}_{cd} \Big\}$.
They  must be  third-kind differentials with poles at $p=t_j, \infty$ (and opposite residues). The residue is read off from \eqref{Npo}. Moreover they  have the property 
\be
 N^{(k)}_{a\ell}\Big\{\Phi_{\ell b}(t_k),N^{(j)}_{cd} \Big\}={\bf 0}, \ \ k\neq j,
\ee
since evaluation at $t_k$ and $N^{(k)} = \v^{(k)} {\h^{(k)}}^t$ Poisson  commute with $N^{(j)} = \v^{(j)} {\h^{(j)}}^t$ and so does $N^{(k)} \Phi(t_k) = \varkappa _k N^{(k)}$.
To see the properties of $N^{(j)} Q(p;c,d)$ for $\ell=j$ we first take into consideration that 
\begin{itemize}
\item $N^{(j)}\Phi(p)$ is analytic at $t_j$;
\item $N^{(j)}\Phi(t_j) =\varkappa _j N^{(j)} \d \zeta_j$;
\item  the evaluation at $t_j$ commutes with the Poisson bracket w.r.t. $N^{(j)}$ (as opposed to a Poisson bracket w.r.t. $\varkappa_j$).
\end{itemize}
From the Leibnitz property of the Poisson bracket we find
\be
N^{(j)}_{a\ell} \Big\{\Phi_{\ell b}(p),N^{(j)}_{cd} \Big\}=  \Big\{N_{a\ell}^{(j)} \Phi_{\ell b}(p),N^{(j)}_{cd} \Big\} 
- \Big\{N_{a\ell}^{(j)} ,N^{(j)}_{cd} \Big\}\Phi_{\ell b}(p) .
\label{13}
\ee
The first term on the right-hand side
is analytic at $p=t_j$ and its evaluation gives $\varkappa _j \d\zeta_j ( \NN jabcd )$ . 
The second  term on the right evaluates to 
\bea
\label{A6}
-\Big\{N_{a\ell}^{(j)} ,N^{(j)}_{cd} \Big\}\Phi_{\ell b}(p)=
-N^{(j)}_{ad} \Phi_{cb}(p) + \delta_{ad}\big( N^{(j)} \Phi(p)\big)_{cb},
\eea
and the last term in \eqref{A6} is analytic at $p=t_j$ with value $\varkappa _j \d \zeta_j N^{(j)}_{cb}\delta_{ad} $, which cancels one term in the right side of the evaluation of \eqref{13}. 
Thus we have 
\be
\label{Qexp}
N^{(j)}_{a\ell} \Big\{\Phi_{\ell b}(p),N^{(j)}_{cd} \Big\} = \varkappa _j \d \zeta_j N^{(j)}_{ad}\delta_{cb} -N^{(j)}_{ad} \Phi_{cb}(p) + \mathcal O(p-t_j).
\ee
Now consider  the expression 
\be
P_{ab;cd}(p):=  \res{q=t_j} \le(\Cauchy(p,q)\Phi(q)\ri)_{ad} \delta_{cb}
 - \res{q=t_j} \Big(\Cauchy_{ad}(p,q)\Phi_{cb}(q)\Big).
 \label{Pabcd}
\ee
The matrices ${\bf P}(p;c,d)$ with entries $P_{ab;cd}(p)$ have the desired residue at $p=t_j$ and are otherwise analytic. Moreover, by the properties of the Cauchy kernel $\Cauchy$ we deduce that  $N^{(\ell)}{\bf P}(t_\ell;c,d)=0$ for $\ell \neq j$.  These $n(ng-1)$ equations  almost determine the holomorphic part. 

To see that they  match the conditions also at $t_j$ we need to multiply by $N^{(j)}$ and look at the regular part as $p\to t_j$. 
Consider the first term in \eqref{Pabcd}: 
\bea
 \res{q=t_j} \le(\Cauchy(p,q)\Phi(q)\ri)_{ad} \delta_{cb}
  =
  \Phi_{ad}(p)\delta_{cb}  
 +
   \res{q=t_j} \le(\Cauchy(t_j,q)\Phi(q)\ri)_{ad} \delta_{cb}
+\mathcal O(p-t_j). 
  \eea
Consider now the second term in \eqref{Pabcd}: it has a pole with singular part equal to that of $\delta_{ad} \Phi_{cb}(p)$:
\bea
\res{q=t_j} \Big(\Cauchy_{ad}(p,q)\Phi_{cb}(q)\Big)  
=
 \delta_{ad} \Phi_{cb}(p) 
 + \res{q=t_j} \Big(\Cauchy_{ad}(t_j,q)\Phi_{cb}(q)\Big) + \mathcal O(p-t_j).
\eea
Thus 
\be
P_{ab;cd}(p) =    \Phi_{ad}(p)\delta_{cb}   - \delta_{ad} \Phi_{cb}(p) 
+   \res{q=t_j} \le(\Cauchy(t_j,q)\Phi(q)\ri)_{ad} \delta_{cb} 
- \res{q=t_j} \Big(\Cauchy_{ad}(t_j,q)\Phi_{cb}(q)\Big) + \mathcal O(p-t_j)
\label{PP}
\ee
After multiplication from the left by $N^{(j)}$ on the left the last two terms in \eqref{PP} vanish and so
\bea
N^{(j)}_{a\ell} P_{\ell b;cd}(p) 
=& \underbrace{\big(N^{(j)}\Phi(p)\big)_{ad}}_{=\varkappa_j \d \zeta_j N_{ad}^{(j)} + \mathcal O(p-t_j)}\delta_{cb}
  - N^{(j)} _{ad} \Phi_{cb}(p) 
+   \mathcal O(p-t_j)=\cr
=&\varkappa_j \d \zeta_j N^{(j)}_{ad} \delta_{cb} - N^{(j)} _{ad} \Phi_{cb}(p)  + \mathcal O(p-t_j)
\eea
and this matches the last requirement \eqref{Qexp}.
\paragraph{Poisson bracket with $\zeta_j$.}
Consider 
\be
\Big\{\Phi(p),\zeta_j\Big\} =
{- \pa_{\varkappa_j} }\Phi(p). 
\ee
This must be a matrix of holomorphic differentials. Since $N^{(j)}\Phi(t_j) = \varkappa _j N^{(j)} \d \zeta_j$ we have 
\be
N^{(j)}\Big\{\Phi(t_j) , \zeta_j\Big\}=\Big\{N^{(j)}\Phi(t_j) , \zeta_j\Big\} =\Big\{\varkappa _j N^{(j)} \d\zeta_j, \zeta_j\Big\}   =
{- N^{(j)} \d \zeta_j}.
\ee
Note, again, that evaluation at $t_j$ commutes with the Poisson bracket with respect to $\zeta_j$. 
Similarly we must have that $N^{(\ell)}\Big\{\Phi(p),\zeta_j\Big\}$ is analytic at $p=t_\ell$ and zero. These are $n^2g$ conditions\footnote{Recall that $N^{(j)} = \v_j \h_j^r$ is of rank one, and so at each Tyurin point while there are $n^2$ equations,  the linearly independent ones are only $n$.} and hence determine the matrix uniquely. By inspection, one sees that 
\be
\Big\{\Phi(p),\zeta_j\Big\} = 
{-  \res{q=t_j} \Cauchy(p,q) \d q_j}.
\ee
Indeed
\be
\lim_{p\to t_j} \res{q=t_j} N^{(j)} \Cauchy(p,q) \d q_j =-   \res{q=p} N^{(j)} \Cauchy(p,q) \d q_j + \res{q=t_j} \lim_{p\to t_j} N^{(j)} \Cauchy(p,q) \d q_j =  N^{(j)}.
\ee
\paragraph{Poisson bracket with $\varkappa _j$.}

Recall  that $\varkappa _j$ is the eigenvalue in the local coordinate $z_j$, namely:
\be
N^{(j)} \frac{\Phi(p)}{\d p_j} \bigg|_{p=t_j} = \varkappa _j N^{(j)}.
\ee
The matrix
\be
\Big\{\Phi(p), \varkappa _j\Big\} =
{ \pa_{\zeta_j} \Phi(p)}
\ee
must have a double pole at $p=t_j$, with singular part proportional to $N^{(j)}$. Indeed, if $\Phi$ is written in the local coordinate $z_j$ then 
\be
\Phi(p) = \le(\frac {N^{(j)}}{p_j-\zeta_j} +\mathcal O(1) \ri)\d p_j
\ee
Then 
\be
\Big\{\Phi(p), \varkappa _j\Big\} =  \le(\frac {N^{(j)}}{( p_j-\zeta_j)^2} + \mathcal O(1)\ri)\d p_j
\ee
The matrix must still satisfy $N^{(\ell)} \Big\{\Phi(p), \varkappa _j\Big\}\bigg|_{p=t_\ell} =0$, for every   $\ell\neq j$. On the other hand for $\ell=j$ we have, differentiating the identity $N^{(j)} \Phi(t_j)=\varkappa _j N^{(j)}\d \zeta_j$ that 
\bea
0 = \pa_{\zeta_j} N^{(j)} \Phi(t_j) =
 N^{(j)} (\pa_{\zeta_j} \Phi)(p)\bigg|_{p=t_j} +  N^{(j)} \d\le(\frac{\Phi(p)}{\d p_j}\ri)\bigg|_{p=t_j}
 =\nn\\
 =
 {N^{(j)}\Big\{\Phi(p), \varkappa _j\Big\}\bigg|_{p=t_j} }+  N^{(j)} \d\le(\frac{\Phi(p)}{\d p_j}\ri)\bigg|_{p=t_j}
\eea
so that the condition becomes
\be
N^{(j)}\Big\{\Phi(p), \varkappa _j\Big\}\bigg|_{p=t_j} =
{-  N^{(j)} \d\le(\frac{\Phi(p)}{\d p_j}\ri)\bigg|_{p=t_j}}
\ee
The following expression satisfies all the requirements
\be
\Big\{\Phi(p), \varkappa _j\Big\} =
{-\res{q=t_j}\Cauchy(p,q)  \d_q \le(\frac{\Phi(q)}{\d q_j}\ri).}
\ee
The proof is complete. \QED

\subsection{Poisson brackets of the kernel with $\Phi$}
Since $\Cauchy$ depends on the moduli of the vector bundle alone, it must Poisson commute with itself:
\be
\Big\{\Cauchy_{ab}(p,q), \Cauchy_{cd}(r,s)\Big\} \equiv 0.
\ee
However it does not commute with $\Phi$. We compute first the Poisson brackets with the various elements of $\Phi$.
\bp
\label{PropKPhi}
\bea
\Big\{\Cauchy_{ab}(q,r) ,N^{(j)}_{cd}\Big\} =&
 - \res{w=t_j}\Cauchy_{ad}(q,w) \Cauchy_{cb}(w,r) 
\label{KN}\\
\label{Kkappa}
\Big\{\Cauchy(q,r) ,\varkappa_j \Big\} = & 
{ \res{w=t_j} \frac{\big(\d_w\Cauchy(q,w)\big)\Cauchy(w,r)}{\d w_j} }
\eea
\ep
\paragraph{Poisson brackets with $N^{(j)}$.}
 As we did before, we observe that $N^{(k)}_{a\ell}\Big\{\Cauchy_{\ell b}(t_k,r) ,N^{(j)}_{cd}\Big\}=0$ for $k\neq j$, while for $k=j$ we have 
\be
N^{(j)}_{a \ell}\Big\{\Cauchy_{\ell b}(q,r) ,N^{(j)}_{cd}\Big\} =\Big\{ \big(N^{(j)}\Cauchy(q,r)\big)_{ab} ,N^{(j)}_{cd}\Big\}
-
\Big\{N^{(j)}_{a \ell} ,N^{(j)}_{cd}\Big\}\Cauchy_{\ell b}(q,r).
\ee
Evaluating at $q=t_j$ sets the first term to zero (recall that evaluation at $t_j$ Poisson commutes with $N^{(j)}$) and so we are left with:
\be
N^{(j)}_{a \ell}\Big\{\Cauchy_{\ell b}(t_j,r) ,N^{(j)}_{cd}\Big\} =-
\Big [ N^{(j)}_{ad}\delta_{c\ell}-\delta_{ad}  N^{(j)}_{c\ell}\Big]\Cauchy_{\ell b}(t_j,r)=
-N^{(j)}_{ad}\Cauchy_{c b}(t_j,r)
\ee
This is solved by 
\be
\Big\{\Cauchy(q,r) ,N^{(j)}_{cd}\Big\} = - \res{w=t_j}\Cauchy_{ad}(q,w) \Cauchy_{cb}(w,r) 
\ee
\paragraph{Poisson brackets with $\varkappa_j$.}
Now we consider
\be
\Big\{\Cauchy(q,r),\varkappa_j\Big\} =
{\pa_{\zeta_j}  \Cauchy(q,r)}.
\ee
Once more multiplication by $N^{(\ell)}$ (on the left)  and evaluation at $q=t_\ell$, $\ell\neq j$ must give zero. It would seem that the same is true for $\ell=j$ but that is not the case because the evaluation at $t_j$ does not commute with the differentiation by $\zeta_j = z_j(t_j)$. So we have 
\be
\le\{
\begin{array}{cc}
N^{(\ell)} \pa_{\zeta_j} \Cauchy(t_\ell,r)= 0 & \ell \neq j\\
N^{(j)} \pa_{\zeta_j} \Cauchy(q,r)\bigg|_{q=t_j}=- N^{(j)}  {\d_q} \le(\frac{\Cauchy(q,r)}{\d q_j}\ri)\Big|_{q=t_j}  .
\end{array}
\ri.
\ee
We solve this interpolation problem as usual and find:
\be
 \pa_{\zeta_j} \Cauchy(q,r)= -\res{w=t_j} \Cauchy(q,w) \d_w \le(\frac{\Cauchy(w,r)}{\d w_j} \ri)=
 \res{w=t_j} \frac{\big(\d_w\Cauchy(q,w)\big)\Cauchy(w,r)}{\d w_j} .
\ee
The proof is complete.
\QED

Combining Proposition \ref{PropKPhi} and Lemma \ref{lemmarecovery} we now prove
\bt
\label{thmKPhi}
\be
\Big\{\Phi_{ab}(w) ,\Cauchy_{cd}(q,r) \Big\}= 
\Cauchy_{ad}(w,r) \Cauchy_{cb}(q,w) 
+ \delta_{cb} \big(\Cauchy(w,q)\Cauchy(q,r)\big)_{ad}
- \Cauchy_{ad}(w,r) \Cauchy_{cb}(q,r)
\ee
\et
\noindent {\bf Proof.}
Using Lemma \ref{lemmarecovery} and the Poisson brackets \eqref{KN}, \eqref{Kkappa} we have 
\bea
\{\Phi_{ab}(w), \Cauchy_{cd}(q,r)\}
=&
\sum \res{x=t_j} \Cauchy_{a\ell}(w,x) \le[\frac{\{N^{(j)}_{\ell b}, \Cauchy_{cd}(q,r)\} }{x_j-\zeta_j} + \{\varkappa_j, \Cauchy_{cd}(q,r)\} \delta_{\ell b} \ri]\d x_j
=\nn
\\
\label{249}
=&\sum \res{x=t_j} \frac{\Cauchy_{a\ell}(w,x)\d x_j}{x_j-\zeta_j} \res{y=t_j}\Cauchy_{c b } (q,y)  \Cauchy_{\ell  d}(y,r)
+
\nn\\
&
{- \sum _j \res{x=t_j} \Cauchy_{a\ell}(w,x)\d x_j \res{y=t_j}\delta_{\ell b} \le( \le(\frac{\d}{\d y_j}\Cauchy(q,y)\ri)\Cauchy(y,r)\ri)_{cd}}
 \eea
We claim that this residue is the same as 
\be
\sum\res{x=t_j} \Cauchy_{cb}(q,x) \big(\Cauchy(w,x) \Cauchy(x,r)\big)_{ad}.
\label{250}
\ee

Once the claim is proved, since we are summing over all residues at $x=t_j$, we can use Cauchy's residue theorem to obtain the opposite residues at $x=q,w,r$, yielding the proof of the theorem. 

To prove the claim we write the expansion of the Cauchy kernel near $t_j$ in the two variables as follows (here $\f = \f^{(j)}$ and $\h=\h^{(j)}$ for brevity):
\bea
\Cauchy_{ab}(u,w) = \frac{ \f_a(u) \h_b}{w_j-\zeta_j} +  \mathbb A_{ab}(u) + \mathcal O(w_j-\zeta_j),\\
\frac{\Cauchy_{ab}(u,w)}{d u_j} = \Cauchy^0_{ab}(w) +\Cauchy^1_{ab}(w) (u_j-\zeta_j) + \mathcal O(u_j-\zeta_j)^2.
\eea
The terms of the expansion should carry an index $_j$ but we have omitted it to keep the notation lighter. 
The matrix $\mathbb A(u)$ is a matrix of one-forms with simple pole at $u=t_j, \infty$ and residue $\1,-\1$, respectively. 
Recall also that $\h_\ell \Cauchy^0_{\ell a}(w) \equiv 0$.
Then the residues at $t_j$ in \eqref{249} yield
\be
\f_c(q) \h_b \mathbb A_{a\ell}(w)\Cauchy^0_{\ell d}(r)  + \f_{a}(w) \h_b  \f_c(q) \h_\ell \Cauchy_{\ell d}^1(r).
\ee
On the other hand the residue at $t_j$ in \eqref{250} gives 
\bea
\res{x_j=\zeta_j}  \le(\frac{ \f_c(q) \h_b}{x_j-\zeta_j} +  \mathbb A_{cb}(q) + \mathcal O(x_j-\zeta_j)\ri)
  \le(\frac{ \f_a(w) \h_\ell}{x_j-\zeta_j} +  \mathbb A_{a\ell}(w) + \mathcal O(x_j-\zeta_j)\ri)
  \le( \Cauchy_{\ell d}^0(r) +\Cauchy_{\ell d}^1(r) (x_j-\zeta_j) + \mathcal O(x_j-\zeta_j)^2\ri)\d x_j
\eea
which simplifies to the same result on account of the fact that $\h_\ell \Cauchy^0_{\ell a}(w) \equiv 0$.
\QED

\bt
\label{thmPhiPhi}
\be
\Big\{\Phi_{ab}(p) ,\Phi_{cd}(q) \Big\}= 
\PhiPhi abpcdq
\ee
\et
{\bf Proof.} We insert the representation \eqref{Phirep} into the Poisson bracket and evaluate the resulting expression term by term using Theorem \ref{thmKPhi} and Proposition \ref{PropPhiPhi}:
\bea
\{\Phi_{ab}(p), \Phi_{cd}(q) \} 
=
\sum_{j}
\le\{ \res{x=t_j} \Cauchy_{a \ell} (p,x) \le[\frac{N^{(j)}_{\ell b}}{x_j-\zeta_j} + \varkappa_j \delta_{\ell b} \ri]\d x_j, \Phi_{cd}(q)\ri\}
=
\nn\\
= \sum_{j}\res{x=t_j} 
\le(\PhiK cdq a\ell px\ri) \le[\frac{N^{(j)}_{\ell b}}{x_j-\zeta_j}  + \varkappa_j \delta_{\ell b} \ri]\d x_j
+
\label{I} \\
-\res{x=t_j} \Cauchy_{a\ell}(p,x)\frac{\d x_j}{x_j-\zeta_j}\le( \PhiN cdq\ell by \ri)
\label{II}\\
{+ \res{x=t_j} \Cauchy_{a\ell}(p,x)\d x_j\delta_{\ell b} \le( \Phikappa cdqjy \ri) 
+\res{x=t_j}  \Cauchy_{a\ell}(p,x) N^{(j)}_{\ell b}  \frac {\d x_j} {(x_j-\zeta_j)^2} \Phizeta cdqjy 
}
\label{III}
\eea
We expand the terms near $x=t_j$ in the local coordinate $z_j$. We set the following notations
\bea
\label{expK}\Cauchy_{ab}(\bullet,x) &= \frac{\f^{(j)}_{a}(\bullet) \h^{(j)}_b} {(x_j-\zeta_j)} + \mathbb A^{(j)}_{ab}(\bullet) + \mathbb B^{(j)}_{ab}(\bullet)(x_j-\zeta_j)  + \mathcal O(x_j-\zeta_j)^2,\\
\label{expPhi}
\frac{\Phi_{ab}(x)}{\d x_j}  &= \frac{N^{(j)}_{ab}}{x_j-\zeta_j} + \Phi_{ab}^{(j),0} + \Phi^{(j),1}_{ab} (x_j-\zeta_j) + \mathcal O(x_j-\zeta_j)^2. 
\eea
Using the representation \eqref{Phirep} for  the second and third term in \eqref{I}, we obtain the new expression
\bea
\eqref{I}=
- \Phi_{cb}(q) & \Cauchy_{ad}(p,q)  
- \delta_{ad}( \Cauchy(q,p) \Phi(p))_{cb} 
\nn\\
+\sum_j&
\Big[ (\mathbb A^{(j)}(q)N^{(j)})_{cb} \mathbb A^{(j)}_{ad}(p) + \f^{(j)}_a(p) \h^{(j)}_d (\mathbb B^{(j)}(q)N^{(j)})_{cb} 
+ \nn\\
&+\varkappa_j \f^{(j)}_c(q)\h^{(j)}_b\mathbb A^{(j)}_{ad}(p) + \varkappa_j \mathbb A^{(j)}_{cb}(q) \f^{(j)}_{a}(p)\h^{(j)}_d \Big]
\eea
Direct evaluation using the expansions \eqref{expK}, \eqref{expPhi}, yields the following expression for the other two lines:
\bea
\eqref{II} =  
-\varkappa_j \mathbb A^{(j)}_{ad}(p)  \f^{(j)}_c(q)\h^{(j)}_b 
-  \mathbb A^{(j)}_{ad}(p)(\mathbb A^{(j)}(q) N^{(j)})_{cb}+\nn\\
+ (\mathbb A^{(j)}(p)N^{(j)})_{ad}\mathbb A^{(j)}_{cb}(q)
+ (\mathbb A^{(j)}(p)\Phi^{(j),0})_{ad} \f^{(j)}_c(q)\h^{(j)}_b.
\eea
\bea
\eqref{III} = 
{
\f_a(p) \h_b \f_c(q) \h_\ell \Phi^1_{\ell d}
- \f_a(p) \h_b( \mathbb B(q) N)_{c d}   
+ (\mathbb B(p) N)_{ ab} \f_c(q)\h_d
}
\eea

Putting them together and using that $N^{(j)}_{ab} = \v^{(j)}_a \h^{(j)}_b$, one verifies that several terms cancel:
\bea
\eqref{I}+ &\eqref{II}+\eqref{III}=
 - \Phi_{cb}(q)  \Cauchy_{ad}(p,q)  
 - \delta_{ad}( \Cauchy(q,p) \Phi(p))_{cb} +
\nn\\
\sum_j \bigg[&
\cancel{ 
(\mathbb A^{(j)}(q)N^{(j)})_{cb} \mathbb A^{(j)}_{ad}(p)} 
+ \cancel{\f^{(j)}_a(p) \h^{(j)}_d (\mathbb B^{(j)}(q)N^{(j)})_{cb} }
+ \nn
\\ 
&+\cancel{\varkappa_j \f^{(j)}_c(q)\h^{(j)}_b\mathbb A^{(j)}_{ad}(p) }
+ \varkappa_j \mathbb A^{(j)}_{cb}(q) \f^{(j)}_{a}(p)\h^{(j)}_d 
+\nn\\
&- \cancel{\varkappa_j \mathbb A^{(j)}_{ad}(p)  \f^{(j)}_c(q)\h^{(j)}_b }
- \cancel{\mathbb A^{(j)}_{ad}(p)(\mathbb A^{(j)}(q) N^{(j)})_{cb} }
\nn\\
&+ (\mathbb A^{(j)}(p)N^{(j)})_{ad}\mathbb A^{(j)}_{cb}(q)
+ (\mathbb A^{(j)}(p)\Phi^{(j),0})_{ad} \f^{(j)}_c(q)\h^{(j)}_b
+\nn
\\
& 
{
\f^{(j)}_a(p) \h^{(j)}_b \f^{(j)}_c(q) \h^{(j)}_\ell \Phi^{(j),1}_{\ell d}
-\cancel{ \f^{(j)}_a(p) \h^{(j)}_b( \mathbb B^{(j)}(q) N^{(j)})_{c d}}   
+ (\mathbb B^{(j)}(p) N^{(j)})_{ ab} \f^{(j)}_c(q)\h^{(j)}_d\bigg].}
\eea
After this simplification we are left with the following expression:
\color{black}
\bea
\eqref{I}+\eqref{II}+\eqref{III}=&
 - \Phi_{cb}(q)  \Cauchy_{ad}(p,q)  
 - \delta_{ad}( \Cauchy(q,p) \Phi(p))_{cb} +
\nn\\
\sum_j\Big[&
 \varkappa_j \mathbb A^{(j)}_{cb}(q) \f^{(j)}_{a}(p)\h^{(j)}_d 
+ (\mathbb A^{(j)}(p)N^{(j)})_{ad}\mathbb A^{(j)}_{cb}(q)
+ (\mathbb A^{(j)}(p)\Phi^{(j),0})_{ad} \f^{(j)}_c(q)\h^{(j)}_b
+\nn
\\
& +\f^{(j)}_a(p) \h^{(j)}_b \f^{(j)}_c(q) \h^{(j)}_\ell \Phi^{(j),1}_{\ell d}
 + (\mathbb B^{(j)}(p) N^{(j)})_{ ab} \f^{(j)}_c(q)\h^{(j)}_d\Big].
 \label{A49}
\eea
Note that the last term in \eqref{A49} is symmetric in the exchange $b\leftrightarrow d$ because $N^{(j)}_{ab} = \v^{(j)}_a \h^{(j)}_b$. 
We claim that the residues in the sum evaluate to 
\be
\sum_{j} \res{x=t_j} \big(\Cauchy(p,x)\Phi(x)\big)_{ad}\Cauchy_{cb}(q,x). 
\label{267}
\ee
Indeed 
\bea
 \res{x=t_j} \big(\Cauchy(p,x)\Phi(x)\big)_{ad}\Cauchy_{cb}(q,x)=\nn
 \\
 = \res{x=t_j} \le(\frac {\f^{(j)}_a(p)\h^{(j)}_\ell }{x_j-\zeta_j} + \mathbb A^{(j)}_{a\ell}(p) +  \mathbb B^{(j)}_{a\ell}(p) (x_j-\zeta_j)+\dots \ri)
\le(\frac {N^{(j)}_{\ell d} }{x_j-\zeta_j} + \Phi^{(j),0}_{\ell d}+   \Phi^{(j),1}_{\ell d}(x_j-\zeta_j)+\dots \ri)\times\nn
\\
\times
 \le(\frac {\f^{(j)}_c(q)\h^{(j)}_b }{x_j-\zeta_j} + \mathbb A^{(j)}_{cb}(q) +  \mathbb B^{(j)}_{cb}(q) (x_j-\zeta_j)+\dots \ri)=\nn
 \\
 =
 \varkappa_j \f^{(j)}_a(p)\h^{(j)}_d\mathbb A^{(j)}_{cb}(q)  
 +  \f^{(j)}_a(p)\h^{(j)}_\ell \Phi^{(j),1}_{\ell d} \f^{(j)}_c(q)\h^{(j)}_b 
 +\big(\mathbb B^{(j)}(p) N^{(j)}\big)_{ad} \f^{(j)}_c(q) \h^{(j)}_b
 +\nn\\
 + \big( \mathbb A^{(j)} (p)\Phi^{(j),0}\big)_{ad} \f^{(j)}_c(q) \h^{(j)}_b  
 +\big(\mathbb A^{(j)}(p)N^{(j)} \big)_{ad} \mathbb A^{(j)}_{cb}(q).
\eea
Applying Cauchy's residue theorem to \eqref{267} yields the result by evaluating the residues at $x=q,p$. \QED

\section{Purely algebraic construction of the Cauchy kernel}
\label{appb}
The formula \eqref{Cauchyentries} seems to require the computation of the  period matrix, which is a transcendental computation. This is unnecessary and one can choose differentials and a third-kind differential that require only algebraic manipulations. 
Fix a non-special divisor $\scr D = \sum_{j=1}^g d_j$. There is a unique (scalar) Cauchy kernel $C(p,q)$ with divisor properties 
\be
\div_p C(p,q) \geq -\infty -q + \scr D, \ \ \ 
\div_q C(p,q) \geq \infty - p - \scr D, \ \ \ \res{p=q} C(p,q)=1.
\ee
Fix local coordinates $z_j$ near $d_j$ and define 
\be
\omega_j(p):= \res{q=d_j} C(p,q) \d z_j(q).
\ee
It is an exercise to see that $\omega_j(p)$ solve a certain interpolation problem and are the higher-genus version of the Lagrange interpolation polynomials. Namely $\omega_j(d_k)=0$ for $j\neq k$ and $\omega_j(p)/\d z_j(p) \to 1$ as $p\to d_j$. One then verifies that the same determinantal formula \eqref{Cauchyentries} works if we substitute $\omega_{q,\infty}(p)$ with this scalar Cauchy kernel $C(p,q)$ and use the $\omega_j$ described here. 

For example if the curve $\CC$ is described by a polynomial equation $E(x,y)=0$ {of the form $y^n+ \dots$}, and $d_j = (x_j,y_j)$ are $g$ generic points satisfying the equation, all the above objects can be written explicitly and require at most inverting matrices. 
For example, if $\infty = (x_\infty, y_\infty)$, $q = (z,w)$ and $p= (x,y)$  then 

\be
C(p,q) = \le(
\frac {E(z,y)} {(x - z )(y-w) } -  \frac { E(x_\infty,y)   } {(x - x_\infty)(y-y_\infty) }  + Q((x,y); (z,w)) \ri)\frac {\d x} {E_{y}(x,y)} 
\ee
where $Q((x,z); (z,w))$ is a polynomial in $x,y$ with powers in the Newton polygon $P$ (thus  corresponding to a holomorphic differential of $p$ and  containing $g$ terms of the form $x^{n_j} y^{m_j}$ with $(n_j,m_j)\in P\subset \N\times \N$, $j=1,\dots g$) obtained by imposing the interpolation  condition that the whole bracket vanishes at $(x,y) = (x_j, y_j)$, $j=1,\dots g$. 
Note also that this $C(p,q)$ is, in fact, the $n=1$ version of the non-abelian Cauchy kernel, and $\scr D$ is the Tyurin divisor for a line bundle. 
\\[10pt]

\noindent{\bf Data availability statement.}
This work did not rely on any additional data. \\[10pt]
\noindent{\bf Conflict of Interest Statement.}
The author declares no conflict of interest.

 \end{document}